\documentstyle[12pt]{article}
\begin{document}
\title{Using Classical Probability To Guarantee Properties of
Infinite Quantum Sequences}
\author{Sam Gutmann \\ Department of Mathematics \\ Northeastern
University \\ Boston, MA  02115 \\ e-mail: sgutm@nuhub.neu.edu }
\date{}
\maketitle
\begin{abstract}
We consider the product of infinitely many copies of a spin-$1\over 2$
system.  We construct projection operators on the corresponding
nonseparable Hilbert space which measure whether the outcome of an
infinite sequence of $\sigma^x$ measurements has any specified property.
In many cases, product states are eigenstates of the projections, and
therefore the result of measuring the property is determined.  Thus we
obtain a nonprobabilistic quantum analogue to the law of large numbers,
the randomness property, and all other familiar almost-sure theorems of
classical probability.
\end{abstract}

By studying infinitely many copies of a quantum system, it is
possible to eliminate references to probability from the postulates of
quantum mechanics.  It suffices to consider an infinite product of
spin-${1\over 2}$ systems.  Statements of the form ``In a
$\widehat{\sigma}^z$-eigenstate, the probability that a single
$\sigma^x$-measurement will yield +1 is $1\over 2$" can be replaced by
``If the state is an infinite product of
$\widehat{\sigma}^z$-eigenstates, half of the
$\sigma^x$-measurements will yield +1" \cite{ref:1}.

Recently Coleman and Lesniewski \cite{ref:2} constructed a
randomness operator, using the classical notion of randomness of a
sequence of +1's and --1's made precise by Kolmogorov and Martin-L\"{o}f.
Their operator measures whether a sequence of independent
$\sigma^x$-measurements yields a random sequence of +1's and
--1's.  If the state is a product of
$\widehat{\sigma}^z$-eigenstates, they show that the answer is {\bf yes}.

In this note, we construct a quantum operator which measures whether
the outcome of the infinite sequence of $\sigma^x$-measurements has
{\bf any} given property.  We use the spectral theorem to provide an
automatic construction of the operator, which is a projection.

Corresponding to each quantum state in the product Hilbert space,
there is a classical probability measure on the set of sequences of
+1's and -1's.  We show how to use this measure to determine how the
operator acts on the state.  For many cases of interest, the classical
probability of the set of sequences which have the property is 0 or 1.
We will show that this implies that the state is an eigenstate of the
operator, with eigenvalue 0 or 1.  That is, if the classical
probability is 1, the state {\bf has} the property.  If the classical
probability is 0, the state doesn't have it.  For example if the state
is a product of $\widehat{\sigma}^z$-eigenstates, the corresponding
measure is the one associated with independent fair coin flips.  If
for sequences generated by this coin flip process the property holds
with probability 1, then the operator will assert that the state has
the property.

Furthermore, suppose the property is a ``tail event", meaning that for
each $n$, it can be described without referring to the first $n$
spins.  Then any product state must be an eigenstate of the
corresponding projection, and again the question of whether the state
has the property has a definite, probability-free answer.

Let $V$ be a two dimensional Hilbert space, and denote the
$\widehat{\sigma}^x$ basis by $|+1\rangle$ and $|-1\rangle$.  To
describe a sequence of independent copies of $V$, we form the
tensor product $V^\infty=V\otimes V\otimes\cdots$, which is a
nonseparable Hilbert space \cite{ref:3}.  The operator
$\widehat{\sigma}^x_n$ corresponds to measuring $\sigma^x$ in the
nth copy; for any $|\psi_j\rangle\epsilon\,V$, with
$\langle\psi_j|\psi_j\rangle=1$, the product state
$|\psi_1\rangle\otimes |\psi_2\rangle\otimes\cdots$ is in $V^\infty$,
and
\begin{eqnarray}
\lefteqn{\widehat{\sigma}^x_n  \left(|\psi_1\rangle\otimes\cdots
|\psi_{n-1}\rangle\otimes |\pm 1\rangle\otimes|\psi_{n+1}
\rangle\otimes\cdots\right)} \\
 &&\;\;\;\;\;\;\; = \pm\left(|\psi_1\rangle\otimes\cdots|\psi_{n-1}
\rangle\otimes|\pm1\rangle\otimes|\psi_{n+1}\rangle\otimes
\cdots\right) \nonumber
\label{eq:1}
\end{eqnarray}
The classical probability analogue to this quantum system is the
(uncountable) set of sequences of +1's and -1's, which we will
denote by $\Omega$.  ``Properties'' of the sequences are subsets of
$\Omega$.  For example, the set of sequences with the (probability
$\frac{1}{2}$) law-of-large-numbers property is
\begin{equation}
\left\{(\sigma_1,\sigma_2,\cdots)\; : \; \lim_{N\rightarrow\infty}
\frac{1}{N} \sum^{N}_{n=1} \sigma_n = 0 \right\}
\label{eq:2}
\end{equation}

Now we use the spectral theorem to construct projections from subsets.  Given
a subset of $\Omega$, let $F(\sigma_1,\sigma_2,\cdots)$
denote its indicator function, i.~e.
$F(\sigma_1,\sigma_2,\cdots)=1$ if $(\sigma_1,\sigma_2\cdots)$ is
in the subset and $F=0$ if not.  The spectral theorem then implies
that the projection operator
\begin{equation}
\widehat{F}=F(\widehat{\sigma}_1^x,\widehat{\sigma}_2^x,\cdots)
\label{eq:3}
\end{equation}
is uniquely defined on all of $V^\infty$.  The nonseparability of
$V^\infty$ is no impediment.  The subsets must be Borel, a condition
that doesn't depend on any particular measure that may be defined on
$\Omega$.  Any subset you can describe (without using the Axiom of
Choice) is Borel.

By using the spectral theorem (and thus taking advantage of the
work done in proving it) we can directly define $\widehat{F}$ by
(\ref{eq:3}), and can avoid considering limits of sequences
of approximations to the desired operators, which are used in
previous approaches \cite{ref:2,ref:4,ref:5}.

On states which are simultaneous $\widehat{\sigma}^x_n$-eigenstates
for all $n$, we have
\begin{equation}
\widehat{F}\left(|\sigma_1\rangle\otimes|\sigma_2\rangle\otimes\cdots\right)
=
F\left(\sigma_1,\sigma_2,\cdots\right)\left(|\sigma_1\rangle\otimes
|\sigma_2\rangle\otimes\cdots\right)
\label{eq:4}
\end{equation}
There are, however, many states in $V^\infty$ which are not
superpositions of these $\widehat{\sigma}_n^x$-eigenstates, and we
want to know how $\widehat{F}$ acts on them.  In particular, for
states of the form
$|\psi_1\rangle\otimes|\psi_2\rangle\otimes\cdots$, which do
constitute an overcomplete basis for $V^\infty$, we want to know
$\|\widehat{F}\left(|\psi_1\rangle\otimes|\psi_2\rangle\cdots\right)\|^2$.

First consider the more familiar case of an operator with a continuous
spectrum, for example, the position operator $\widehat{x}$.  With
$\widehat{G}=G(\widehat{x})$ we can write
\begin{equation}
\|\widehat{G}|\psi\rangle\|^2 = \int |G(x)|^2 \;\; |\langle
x|\psi\rangle|^2 dx
\label{eq:5}
\end{equation}
Note that the measure given by $d\mu = |\langle x|\psi\rangle|^2 dx$
depends on $|\psi\rangle$ but not on $G$.

Similarly, with
$\widehat{G}=G\left(\widehat{\sigma}^x_1,\widehat{\sigma}^x_2,\cdots\right)$
we have
\begin{equation}
\|\widehat{G}(|\psi_1\rangle\otimes|\psi_2\rangle\otimes\cdots)\|^2 = \int
| G(\sigma_1,\sigma_2,\cdots)|^2 d\mu (\sigma_1,\sigma_2,\cdots)
\label{eq:6}
\end{equation}
for any bounded Borel-measurable $G$.
(The cases of (\ref{eq:5}) and (\ref{eq:6})
are not completely analogous.  In (\ref{eq:5}), as
$G$ varies, $\widehat{G}|\psi\rangle$ can span all of the Hilbert space.
In (\ref{eq:6}), the span of
$\widehat{G}(|\psi_1\rangle\otimes|\psi_2\rangle\otimes\cdots)$ can only be a
(separable) subspace of $V^\infty$ called the component of
$|\psi_1\rangle\otimes|\psi_2\rangle\otimes\cdots)$.

As in (\ref{eq:5}), the measure $\mu$ in (\ref{eq:6}) depends on
$|\psi_1\rangle\otimes|\psi_2\rangle\otimes\cdots$ but not on $G$.
We'll now determine $\mu$ in terms of
$|\psi_1\rangle\otimes|\psi_2\rangle\otimes\cdots$ by choosing $G$ to be
indicator functions depending only on finitely many
$\widehat{\sigma}^x_n$.  For indicator functions $F$, formula
(\ref{eq:6}) becomes
\begin{equation}
\|\widehat{F}\left(|\psi_1\rangle\otimes|\psi_2\rangle\otimes\cdots
\right) \|^2 = \int F\left(\sigma_1,\sigma_2,\cdots\right)
d\mu (\sigma_1,\sigma_2,\cdots)
\label{eq:7}
\end{equation}
First consider the case of one
factor of $V$.  Let
$\langle\psi_1|\psi_1\rangle=1$.
As an example, let $f(\sigma_1)=1$ if $\sigma_1=+1$ and $0$ if
$\sigma_1=-1$.  Then $\|\widehat{f}|\psi_1\rangle\|^2 =
\|f\left(\widehat{\sigma}_1^x\right)|\psi_1\rangle\|^2 = \|\langle
+1|\psi_1\rangle\|^2$.  For any of the four possible indicator
functions $f$ we can write
\begin{equation}
\|\widehat{f}|\psi_1\rangle\|^2 = \int f(\sigma_1) d\mu
(\sigma_1)
\label{eq:8}
\end{equation}
where the measure $\mu$ assigns point masses to +1 and --1 by
$\mu(+1)=|\langle +1|\psi_1\rangle|^2 = 1-|\langle
-1|\psi_1\rangle|^2 = 1-\mu(-1)$.  This follows from how
$f\left(\widehat{\sigma}_1^x\right)$ acts on eigenstates.

Similarly in the $V^\infty$ case, given a state
$|\psi_1\rangle\otimes|\psi_2\rangle\otimes\cdots$ with each
$\langle\psi_j|\psi_j\rangle=1$, let $|\langle +1|\psi_j\rangle|^2 =
p_j$.  (This definition of $p_j$ makes explicit that the measure depends
on the state $|\psi_1\rangle\otimes |\psi_2\rangle\otimes\cdots$ and on
the decision to measure $\sigma^x$.)  Let $\mu$ be the measure on
$\Omega$ which assigns probability $p_j$ to the set of sequences with
$\sigma_j=+1$, and for which the coordinates
$\sigma_1,\sigma_2,\cdots\sigma_n$ are independent for any $n$, i.~e.
\begin{eqnarray}
\lefteqn{\mu\left\{\left(\sigma_1,\sigma_2,\cdots\right):
\sigma_1=\omega_1,\sigma_2=\omega_2,\cdots,\sigma_n=\omega_n\right\}}
\label{eq:9}
\\  && = p_1^{\omega_1}(1-p_1)^{1-\omega_1}\cdots p_n^{\omega_n}
(1-p_n)^{1-\omega_n} \nonumber
\end{eqnarray}
This expression for $\mu$ follows from repeating the
one-dimensional argument for $n$ dimensional functions.
Probability theory ensures that $\mu$ is completely determined by
(\ref{eq:9}).

We are now equipped to connect directly classical and quantum
statements.  For example, consider the (probability $\frac{1}{2}$)
law-of-large-numbers property: $F(\sigma_1,\sigma_2,\cdots)=1$ if
$\lim_{N\rightarrow\infty}\frac{1}{N}\sum^{N}_{n=1}\sigma_n=0$ and
$F=0$ if not.  If $|\psi_n\rangle$ is a
$\widehat{\sigma}^z$-eigenstate for each $n$, say $|\psi_n\rangle =
\frac{1}{\sqrt{2}} |+1\rangle + \frac{1}{\sqrt{2}} |-1\rangle$, then
$p_n=\frac{1}{2}$ for each $n$.  The classical strong law of large
numbers asserts $\int
F(\sigma_1,\sigma_2,\cdots)d\mu(\sigma_1,\sigma_2, \cdots)=1$ and by
(\ref{eq:7}) we have
$\|\widehat{F}\left(|\psi_1\rangle\otimes|\psi_2\otimes\cdots\right)\|^2
=1$, implying
\begin{equation}
\widehat{F}\left(|\psi_1\rangle\otimes|\psi_2\rangle\otimes\cdots\right)
= |\psi_1\rangle\otimes|\psi_2\rangle\otimes\cdots
\label{eq:10}
\end{equation}
Thus if we measure the law-of-large-numbers projector on this product of
$\widehat{\sigma}^z$-eigenstates, we obtain 1.

Randomness works the same way.  $F(\sigma_1,\sigma_2,\cdots)=1$ if
$(\sigma_1,\sigma_2,\cdots)$ has the Kolmogorov-Martin-L\"{o}f
randomness property.  Use the same $|\psi_n\rangle$ as above, so $p_n$
and $\mu$ remain the same.  Again, classical probability says that
according to $\mu$, almost every sequence $(\sigma_1,\sigma_2,\cdots)$
is random \cite{ref:6}.  So
$\widehat{F}(|\psi_1\rangle\otimes|\psi_2\rangle\otimes\cdots) =
|\psi_1\rangle\otimes |\psi_2\rangle\otimes\cdots$ where now
$\widehat{F}$ is the randomness operator.  Of course, the same
conclusion holds for any property which has probability 1 according to
$\mu$.  Analogous results hold for different $|\psi_1\rangle\otimes
|\psi_2\rangle\otimes\cdots$ which give rise to different $\mu$.

Properties such as the law of large numbers and randomness are tail
events; that is, for each $n$, they depend only on
$\sigma_{n+1},\sigma_{n+2},\cdots$ (Clearly
$\lim_{N\rightarrow\infty}\frac{1}{N}\sum^{N}_{m=1}\sigma_m =
\lim_{N\rightarrow\infty}\frac{1}{N}\sum^{\infty}_{m=n+1}\sigma_m$.  See
\cite{ref:6} for details about randomness that show it is a tail
event.)  The classical Kolmogorov zero-one law \cite{ref:7} asserts that
if $F$ is the indicator function of a tail event, and $\mu$ is
determined by $p_1,p_2,\cdots$ as in (\ref{eq:7}), then
\begin{equation}
\int F\left(\sigma_1,\sigma_2,\cdots\right)
d\mu\left(\sigma_1,\sigma_2,\cdots\right)=0 \,
\hbox{or}\, 1
\label{eq:11}
\end{equation}
Thus for any $|\psi_1\rangle\otimes|\psi_2\rangle\otimes\cdots$ and $F$
corresponding to a tail event,
$|\psi_1\rangle\otimes|\psi_2\rangle\otimes\cdots$ must be an eigenstate
of the projection $\widehat{F}$.  Every product state
$|\psi_1\rangle\otimes|\psi_2\rangle\otimes\cdots$ either {\bf has} or
{\bf doesn't have} any given tail property.

So product states are eigenstates of projection operators corresponding
to many interesting properties.  The derivation of this fact and the
calculation of the eigenvalue requires classical probability theory, but
the quantum statement makes no reference to probability.
\newpage
\noindent
{\Large\bf Acknowledgements}
\vspace{2ex}

Thanks to S.~Coleman and A.~Lesniewski for helpful discussions, to
E.~Farhi for trying to lengthen this paper, to J.~Goldstone for trying
to shorten it, and to C.~Lewis for infinite patience.

\end{document}